\documentclass[12pt]{article}
\usepackage{lscape}
\usepackage{graphicx}

\begin{document}

\title{ Vortex solitons in  dispersive nonlinear Kerr type media}

\author{L. M. Kovachev, L. M. Ivanov, \\
D. Y. Dakova$^\dag$, L.I. Pavlov$^{\dag\dag}$, K. L. Kovachev,  \\
Institute of Electronics, Bulgarian Academy of Sciences,\\
Tsarigradcko chaussee 72,1784 Sofia, Bulgaria,\\
$^\dag$ Plovdiv Univetsity, 24 Tsar Asen, 400, Plovdiv, Bulgaria\\
$^{\dag\dag}$ Institute for Nuclear Research and Nuclear Energy,\\
72 Tsarigradsko chaussee,1784, Bulgaria,} \maketitle
\date{}
\begin{abstract}

We have investigated the nonlinear amplitude vector equation
governing the evolution of optical pulses in optical and UV region.
We are normalizing this equation for the cases of different and
equal transverse and longitudinal size of optical pulses or
modulated optical waves, of week and strong dispersion. This gives
us the possibility to reduce the amplitude equation to different
nonlinear evolution equations in the partial cases. For some of
these nonlinear  equations exact vortex solutions are found.
Conditions for experimental observations of these vortices are
determined.  We have found large spectral zones on the 'wings' of
the electron resonances of metal vapors and also near to the plasma
frequency where  the amplitude equation can be reduced to 3D+1
vector nonlinear Schr\"odinger equation.

PACS 42.81.Dp;05.45.Yv;42.65.Tg
\end{abstract}

\section{Introduction}
The investigation of optical pulses and  of modulated periodical
waves in wave guide structures as fibers and planar waveguides is
connected with the additional requirements to the optical property
of the materials. One standard way for confinement the optical beams
and pulses is by variation of the linear or nonlinear refractive
index in plane, which is transverse to the propagation direction. In
an isotropic dispersive media the refractive index depends only on
the frequency, the group velocity depends on the chromatic
dispersion and the wave guide modes can not be introduced in linear
regime of propagation. To provide one correct analysis in this case
we need to get in mind also the different spatial size of the
optical pulses. The longitudinal spatial size is determined from the
time duration by the relations $z_0=v_gt_0=ct_0/n_g$, where a
typical value of the group index $n_g$ in transparency region of
pure silica is ($n_g\cong 1.5$). For example, the typical transverse
size  of  laser pulses is evaluated from  $r_{\perp}=1-5 $ $mm$ to
$r_{\perp} =100$ $\mu m$. While for the longitudinal size of pulses
or spatial period of modulated waves, there are mainly two
possibilities: a. Pulses from nanoseconds up to $40-100$
picoseconds. For such pulses $z_0>>r_{\perp}$ and their shape is
more close to  optical filaments. b. Pulses from few picoseconds to
$100$ femtoseconds. In this case $z_0\sim r_{\perp}$ and the pulses
look like as optical bullets. When the transverse and longitudinal
part are approximately equal (case b), the ignoring of the second
derivative in $z$ direction in amplitude equation is not possible.
The propagation of such type of short optical pulses in nonlinear
media must be characterized, not only with the equal dimensions in
$x$, $y$ and $z$ direction, but also with non-stationary optical
linear and nonlinear response of the media. That is why we can
expect, that the equations governing the propagation of such type of
pulses, like optical bullets, are different from the well known
paraxial approximated equations, which describe evolution of
nanosecond and picosecond pulses \cite{CHIAO,TAL,KEL}. Another basic
characteristic of the pulses is their polarization. As it is well
known, in the case of plane monochromatic wave, the electromagnetic
field is polarized in the plane perpendicular to the direction of
propagation. In this case four Stokes parameters, two amplitudes and
two phases, connected with the transverse components of the
electromagnetic field can be introduced \cite{BORN}. This
corresponds to SU(2) symmetry of propagation of the monochromatic
field. The nanosecond and hundred picosecond optical pulses admit
very little spectral bandwidth, and the polarization component in
the propagation direction is small in respect to the transverse
components. With good approximation these pulses or modulated waves
also can be investigated as polarized in $x,y$ plane. Their spatial
form also is  more close to the cw wave, as the longitudinal size is
hundred times greater, than the transverse size.   In contrast to
the case of long pulses, the few ps and femtosecond optical waves
are with relatively equal size in $x,y,z$ direction and admit large
spectral width. For them such coherent SU(2) tensor can not be
introduced. The experimental results show, that the electric field
of sub-picosecond and femtosecond pulses, always contains also third
longitudinal component. With the enlargement of the spectral width
of the vector field also the longitudinal component grows (the
component normal to the standard Stokes coherent polarization
plane). Recently Carozzi et al. have shown that for the wave packets
higher class of symmetry - SU(3) exists \cite{BERG}. Using the
presentation of Gell-Mann of SU(3), they prove that five independent
parameters, three amplitudes and two phases, define the dynamics of
propagation of the vector field. This corresponds to a
three-component vector field. The above made analysis and the
experimental results show that in the investigation of
sub-picosecond and femtosecond pulses we must always have in mind
the three-component vector character of the electromagnetic field.
An initial investigation of the propagation of the optical pulses
was provided in the frame of spatiotemporal scalar evolution
equation \cite{SIL,DRA}. We found that vector generalization of such
spatiotemporal model can be applied to optical pulses or to
modulated periodical waves in UV region for dielectrics with strong
dispersion.

\section{Nonlinear amplitude vector equation}

Starting from the Maxwell equation for an isotropic, dispersive,
nonlinear Kerr-type media with no stationary response, we derive
the next slowly varying vector amplitude equation \cite{KOV}:

\begin{eqnarray}
\label{a11} i\left(\frac{\partial\vec A}{\partial t} +
v\frac{\partial\vec A}{\partial z}+\left(n_2+\frac {
k_0v}{2}\frac{\partial n_2}{\partial\omega}\right)\frac{\partial
\left(\left|\vec A\right|^2\vec
A\right)}{\partial t} \right) =\nonumber\\
\\
\frac{v}{2k_0}\Delta\vec A -
\frac{v}{2}\left(k_0^{"}+\frac{1}{k_0v^2}\right)
\frac{\partial^2\vec A}{\partial t^2} + \frac{n_2 k_0
v}{2}\left|\vec A\right|^2\vec A,\nonumber
\end{eqnarray}
where $k_0$, $v$, $k_0^{"}$ and $n_2$ are the carrying wave number,
group velocity, dispersion of the group velocity, and the nonlinear
refractive index respectively. With $\vec A $ we denote the slowly
varying vector amplitude of the electrical field. The equation
(\ref{a11}) is written in second approximation to the linear
dispersion and in first approximation to the nonlinear dispersion.
The kind of nonlinearity is connected with the fact, that in this
paper, as in the previous one \cite{KOV}, we investigate  only
linearly polarized electrical field. The dependance of the nonlinear
polarization from  different kinds of polarizations of the
electrical field is discussed in the Appendix 1. We will apply to
the basic amplitude equation (\ref{a11}) the method of different and
equal transverse and longitudinal spatial scales, as it is connected
with the natural initial shape of the optical pulses. This
application is one simplification of the multi-scale method,
introduced in the nonlinear optics in \cite{ACX, ASC, NEW} and also
in the hydrodynamics \cite{NL}.

\section{Case of different transverse and longitudinal size}
Applying a "moving in time" ($t' = t-z/v; z' = z $) transformation
to the vector amplitude equation (\ref{a11}), we obtain:

\begin{eqnarray}
\label {mt} -i\left(\frac{\partial\vec A}{\partial z'}+
\frac{1}{v}\left(n_2+\frac { k_0v}{2}\frac{\partial
n_2}{\partial\omega}\right)\frac{\partial \left(\left|\vec
A\right|^2\vec A\right)}{\partial t'} \right)+
\frac{1}{2k_0}\Delta_{\bot}\vec A -
\frac{k_0^{"}}{2}\frac{\partial^2\vec A}{\partial t'^2}
-\nonumber\\
\\
 \frac{1}{2k_0}
\left(\frac{\partial^2\vec A}{\partial
z'^2}-\frac{2}{v}\frac{\partial^2\vec A}{\partial t'\partial
z'}\right)+ \frac{n_2 k_0 v}{2}\left|\vec A\right|^2\vec
A=0,\nonumber
\end{eqnarray}
where $\Delta_{\bot}=\frac{\partial^2}{\partial x^2} +
\frac{\partial ^2}{\partial y^2}$.
The estimation of the influence of different terms in equation (\ref{mt})
can be reached by  writing it in dimensionless variables. To make one assessment of
the difference between transverse and longitudinal dimension of the pulses we are introducing separately
transverse $r_\perp$ and longitudinal $z_0$ constants. Defining the rescaled
variables:

\begin{eqnarray}
\label{eq12} \vec A=A_0\vec A';\ x=r_{\perp}x';\ y=r_{\perp}y';\
z'=z_0z";\ t'=t_0t",
\end{eqnarray}
 and constants:

\begin{eqnarray}
\label{norm}
z_0/t_0=v;\ \alpha=k_0z_0;\ \delta=r_{\perp}/z_0;\ z_{dis}=t_0^2/k";\ \nonumber\\
\\
\beta_1=k_0r_{\perp}^2/z_{dis};\
\gamma=k_0^2r_{\perp}^2n_2\left|A_0\right|^2/2;
\gamma_1=n_2\left|A_0\right|^2/2, \nonumber
\end{eqnarray}
equation (\ref{mt}) can be transformed in the following (the
primes and the seconds are not written for clarity):

\begin{eqnarray}
\label{tp}
 -i\alpha\delta^2\left(\frac{\partial\vec A}{\partial z}+
\gamma_1\frac{\partial \left(\left|\vec A\right|^2\vec
A\right)}{\partial t} \right)+\frac{1}{2} \Delta_{\bot}\vec A
-\frac{1}{2}\beta_1\frac{\partial^2\vec A}{\partial t^2}
-\nonumber\\
\\
\frac{1}{2}\delta^2\left(\frac{\partial^2\vec A}{\partial
z^2}-2\frac{\partial^2\vec A}{\partial t\partial z}\right) +
\gamma\left|\vec A\right|^2\vec A=0, \nonumber
\end{eqnarray}
where $z_{dis}=t_0^2/k"$ and in the transparency region we use the
approximation \\ $\partial n_2/\partial\omega\simeq 0$.

\subsection{Pulse propagation in optical region: week dispersion}
In the optical region the wave vector is valued from $k_0\sim 10^4$
to $k_0\sim 10^5$ $cm^{-1}$. The typical value of the dispersion
parameter in the transparency optical region of dielectrics is from
$k"\sim 10^{-28}$ $s^2/cm$  up to $k"\sim 10^{-24}$ $s^2/cm$ in UV
region \cite{AXM}. The transfer sizes $r_{\perp}$ of laser generated
optical pulses or modulated periodical in time waves are valued from
$r_{\perp}< 1 $ to $r_{\perp}\sim10^{-2}$ $cm$. We will investigate
laser pulses or modulated periodical waves with time duration or
modulation period from few nanoseconds ($t_0\sim 10^{-9}$ $s$) up to
$40-100$ picoseconds ($t_0\sim 4-10 \times 10^{-11}$ $s$) in the
transparency region of a nonlinear Kerr type media. The group
velocity index in this case is about $n_g\simeq1.5$ and using the
relations $z_0=v_gt_0=ct_0/n_g\sim 10^{1}-10^{2}$ $cm$, we find that
the dimensionless parameter in  front of second derivative in $z$
direction and cross-term is very small
$\delta^2=r_{\perp}^2/z_0^2\simeq 10^{-2}-10^{-4}$. Another
important relation is that if slowly-varying amplitudes are used
$\alpha=k_0z_0\simeq 10^{3}-10^{4}$. For such typical values of the
dispersion $k"$,  wave vector $k_0$, time duration $t_0$ and
transverse dimension $r_{\perp}$ of the optical wave, the
dimensionless parameter $\beta_1$ in front of the dispersion term in
equation (\ref{tp}) is very small  $\beta=k_0r_{\perp}^2/z_{dis}\sim
10^{-3}-10^{-4}$ . Here we study the cases with power near the
critical for self-focusing $\gamma \cong 1$. These valuations give
us possibility to estimate dimensionless constant in front of the
different terms of normalized amplitude equation (\ref{tp}) for the
transparency optical region of a dispersive nonlinear Kerr type
media:

\begin{eqnarray}
\label{norm1}
\alpha=k_0z_0\simeq 10^{3}-10^{4};\
\delta^2=r_{\perp}^2/z_0^2\simeq 10^{-2}-10^{-4};\
\nonumber\\ \beta_1=k_0r_{\perp}^2/z_{dis}\simeq 10^{-3}-10^{-4};\ \\
\gamma=k_0^2r_{\perp}^2n_2\left|A_0\right|^2\simeq 1;\
\gamma_1=n_2\left|A_0\right|^2\ll 1. \nonumber
\end{eqnarray}
Neglecting the small terms in (\ref{tp}) we obtain the next amplitude equation:

\begin{eqnarray}
\label{parax}
 -i\frac{\partial\vec A}{\partial z}+
\frac{1}{2} \Delta_{\bot}\vec A +
\left|\vec A\right|^2\vec A=0.
\end{eqnarray}
As it can be expected, the dynamics of such kind of long optical
pulses is governed by the well known paraxial approximation. This
equation was introduced initially for optical beam
\cite{CHIAO,TAL,KEL}. The fact, that with such type  of  nonlinear
paraxial equation it is possible to investigate non-stationary
processes is well known also \cite{CHIR,LIT,ASC}.  It seems natural
the dynamics of propagation of long pulses to be governed by the
equation which describes dynamics of optical beam. The paraxial
approximation is applicable even when the relation between the
longitudinal and the transverse part is of the order of ten
($\delta=r_{\perp}/z_0\sim 1/10$). The dimensionless coefficient in
front of the term with second derivative in z direction and
cross-term is of the order of square of $\delta$ ( $\delta^2\sim
1/100<<1$), the term can be neglected and this leads again to the
paraxial equation (\ref{parax}). Effects of self-steepening due to
the first order of the nonlinear dispersion (second term in equation
(\ref{tp}))  is not possible as always $\gamma_1\approx
n_2|A|^2<<1$.  Different kind of generalization of equation
(\ref{parax}) in respect to nonlinearities \cite{SIL,KIV,DESM,TOR}
and multi-component vector fields \cite{MAN,DES} was performed. It
is important to point here that the second derivative in $z$
direction and the crossed term are neglected from the difference
between transverse and longitudinal size
($\delta^2=r_{\perp}^2/z_0^2\simeq 10^{-2}-10^{-4}<<1$), and not
because the slowly varying amplitude is used. When we investigate
the optical pulses with equal transverse and longitudinal size
$\delta^2=r_{\perp}^2/z_0^2\simeq 1$, the second derivative in $z$
direction and cross term of the amplitude function are of the same
order that the transverse Laplassian, and these terms influence
considerably on the dynamics of propagation. This important case we
will investigate in the next section. When the power of the pulse is
less than  the critical for self-focusing $\gamma << 1$ the
diffraction term dominates  and the propagation of the optical
pulses is governed by the next linear paraxial equation:

 \begin{eqnarray}
\label{paraxl}
 -i\frac{\partial\vec A}{\partial z}+
\frac{1}{2} \Delta_{\bot}\vec A=0.
\end{eqnarray}

\subsection{Optical pulses in UV region: strong dispersion. Vortex solutions}

In UV region of media with high density as liquids and  dielectrics
the dispersion  $k"$ increases considerably and can reach the values
from  $k"\sim 10^{-24}-10^{-25}$ $s^2/cm$. The carrying wavenumber
is valued from $k_0\sim 10^5$ $cm^{-1}$ to $k_0\sim 10^6$ $cm^{-1}$
and in this spectral region we can reach value with equal
diffraction and dispersion length, which corresponds to $
\beta_1=k_0r_{\perp}^2/z_{dis}\sim 1$.   We use for example  bulk
fused  silica and calculate the linear refractive index from the
Sellmeier relations \cite{MAL}:

\begin{eqnarray}
\label{Sel}
n^2=1+\sum_{j=1}^{3}{\frac{B_j\omega_j}{\omega_j-\omega}},
\end{eqnarray}
with coefficients $B_1=0.6961663,\ B_2=0.4079426,\ B_3=0.897479$,
wavelengths $\lambda_1=0.0684043$ $\mu m$ $\lambda_2=0.1162414$ $\mu
m$ $\lambda_3=9.896161$ $\mu m$ where $\lambda_j=2\pi c/\omega_j$.
The  choice of  carrying wavelength for the optical packet at
$\lambda=0.264$ $\mu m$, ($\omega=0.7131606. 10^{16}$  $Hz$) is
connected with the facts, that fused silica is still transparent on
this wavelength and it can be reached easy way by fourth harmonics
from Nd YAG laser on $\lambda=1.064$ $\mu m$. Considering the case
of pulses or modulated waves with intensity near the critical for
self-focusing $\gamma\sim1$, time duration $t_0\simeq 30-40$ $ps$
and using the relations $k(\omega)=\sqrt{\omega^2n^2(\omega)/c^2},\
k"=\frac{\partial^2 k}{\partial \omega^2} $, the next values for
laser characteristics and material constants are obtained:
\begin{center}
\begin{eqnarray}
\label{norm2}
k_0=2.378\times 10^5\ cm^{-1};\  k"=1.99\times 10^{-25}\ s^2/cm;\ n_g\sim 1.65;\ \nonumber\\
r_\perp\simeq (1.4-1.6)\times 10^{-1}\ cm;\
z_{dis}\simeq (0.45-0.8)\times 10^{4}\ cm;\ \\
z_0\sim (6-8)\times 10^{-1}\ cm;\ k_0^2r_\perp^2\simeq
(1.1-1.5)\times 10^{9};\ \nonumber\\
 n_2|A_0|^2\simeq (1.1-1.5)\times 10^{-9}. \nonumber
\end{eqnarray}
\end{center}
These parameters correspond to the next dimensionless constant:

\begin{eqnarray}
\label{norm3}
\beta_1=k_0r_{\perp}^2/z_{dis}\simeq 1;\
\alpha\cong 10^{4};\ \nonumber\\
\delta^2=r_{\perp}^2/z_0^2\simeq 10^{-2};
\gamma=k_0^2r_{\perp}^2n_2\left|A_0\right|^2\simeq 1; \\
\alpha\delta^2\gamma_1=\alpha \delta^2n_2\left|A_0\right|^2\simeq
10^{-6}\ll 1. \nonumber
\end{eqnarray}
Neglecting the small terms from the amplitude equation (\ref{tp}) we
obtain the next equation governing the propagation of optical pulses
with different transfer and longitudinal size in UV region of
dielectrics:

\begin{eqnarray}
\label{UV}
 -i\alpha\delta^2\frac{\partial\vec A}{\partial z}+
\frac{1}{2} \Delta_{\bot}\vec A
-\frac{1}{2}\beta_1\frac{\partial^2\vec A}{\partial t^2}
+\gamma\left|\vec A\right|^2\vec A=0.
\end{eqnarray}
The scalar variant of this spatiotemporal equation (\ref{UV}) was
introduced for first time by Silberberg \cite{SIL} and attracts a
growing interest. In the beginning this equation was suggested to
discover the dynamics of so-called "light bullets" (LB), i.e.
optical pulses with relatively equal transfer and longitudinal size.
Our investigation corrects this idea and finds that the equation
(\ref{UV}) governed the propagation of optical pulses with
longitudinal size hundred times greater than the transfer one in UV
region of dielectrics and liquids. In the case of optical waves with
very small intensity the nonlinear term can be neglected
$\gamma<<1$, and we obtain the linear version of the  equation
(\ref{UV}):

\begin{eqnarray}
\label{UVL}
 -i\alpha\delta^2\frac{\partial\vec A}{\partial z}+
\frac{1}{2} \Delta_{\bot}\vec A
-\frac{1}{2}\beta_1\frac{\partial^2\vec A}{\partial t^2}
=0.
\end{eqnarray}
Let investigate the two component vector field polarizable in
transverse $x,y$ plane. For long pulses, as they are more close to
the cw regime, the longitudinal vector component is small and such
approximation is possible. We represent the optical vector field by
two orthogonal components $A_1\vec{x}+ A_2\vec{y}$ and the equation
(\ref{UV}) is transformed to the next nonlinear system of equations:

\begin{eqnarray}
\label{SYSUV}
 -i\alpha\delta^2\frac{\partial A_1}{\partial z}+
\frac{1}{2} \Delta_{\bot} A_1
-\frac{1}{2}\beta_1\frac{\partial^2 A_1}{\partial t^2}
+\gamma \left(| A_1|^2+ | A_2|^2\right)A_1=0, \nonumber\\
\\
-i\alpha\delta^2\frac{\partial A_2}{\partial z}+
\frac{1}{2} \Delta_{\bot} A_2
-\frac{1}{2}\beta_1\frac{\partial^2 A_2}{\partial t^2}
+\gamma \left(| A_2|^2+ | A_1|^2\right)A_2=0. \nonumber
\end{eqnarray}
In the linear case the (\ref{SYSUV}) is reduced to two equal scalar equations for $A_1$ and $A_2$:

\begin{eqnarray}
\label{LSUV}
 -i\alpha\delta^2\frac{\partial A_i}{\partial z}+
\frac{1}{2} \Delta_{\bot} A_i
-\frac{1}{2}\beta_1\frac{\partial^2 A_i}{\partial t^2}=0,
\end{eqnarray}
where $i=1,2$.
We rewrite the linear scalar equation (\ref{LSUV}) in cylindrical coordinates:

\begin{eqnarray}
\label{LC}
-i\alpha\delta^2\frac{\partial A_i}{\partial z}+
\frac{1}{2}\left( \frac{1}{r}\frac{\partial A_i}{\partial r}+
\frac{\partial^2 A_i}{\partial r^2}+
\frac{1}{r^2}\frac{\partial^2 A_i}{\partial \theta^2} \right)
-\frac{1}{2}\beta_1\frac{\partial^2 A_i}{\partial t^2}=0,
\end{eqnarray}
where $r=\sqrt{x^2+y^2}$  and $\theta=\arctan(x/y)$.
The linear equation (\ref{LC}) admits infinite number of exact analytical solutions of kind of:

\begin{eqnarray}
\label{LSOL}
A_i=r^{n}\left[a_n\cos(n\theta)+a_n\sin(n\theta)\right]\exp\left(i(\beta_1
z+\sqrt{2\alpha\delta^2}t)\right),
\end{eqnarray}
\begin{eqnarray}
\label{LSOL1}
A_i=r^{-n}\left[a_n\cos(n\theta)+a_n\sin(n\theta)\right]\exp\left(i(\beta_1
z+\sqrt{2\alpha\delta^2}t)\right),
\end{eqnarray}
where $n=0,1,2...\infty$ is one infinite number. The solutions
(\ref{LSOL}), (\ref{LSOL1})  admit modulation frequency
$\Omega=\sqrt{2\alpha\delta^2}$ and modulation wave vector
$K=\beta_1$ in z direction. The existence of such kind solutions for
monochromatic light was observed for first time by Nye and Berry
\cite{NB}. The solutions of kind  (\ref{LSOL}),(\ref{LSOL1}) admit
amplitude and phase singularities. One elegant way to remove the
amplitude singularities by using the combination of Gaussian
envelope and the solution (\ref{LSOL1}) with $n=1$ was suggested for
first time in \cite{SOS} for paraxial beam. This method is
applicable also to the linear equation (\ref{LC}) and optical
vortices in combined pulses with finite amplitudes can be found.
Usually the localized solutions of the linear equations are not
stable. Now we will turn to most interesting case; exact solutions
of the nonlinear system of equation (\ref{SYSUV}). Again we rewrite
the system (\ref{SYSUV}) in cylindrical variables:

\begin{eqnarray}
\label{NLC}
 -i\alpha\delta^2\frac{\partial A_1}{\partial z}+
\frac{1}{2}\left( \frac{1}{r}\frac{\partial A_1}{\partial r}+
\frac{\partial^2 A_1}{\partial r^2}+ \frac{1}{r^2}\frac{\partial^2
A_1}{\partial \theta^2} \right)
\nonumber\\-\frac{1}{2}\beta_1\frac{\partial^2 A_1}{\partial t^2}
+\gamma \left(| A_1|^2+ | A_2|^2\right)A_1=0, \nonumber\\
\\
 -i\alpha\delta^2\frac{\partial A_2}{\partial z}+
\frac{1}{2}\left( \frac{1}{r}\frac{\partial A_2}{\partial r}+
\frac{\partial^2 A_2}{\partial r^2}+ \frac{1}{r^2}\frac{\partial^2
A_2}{\partial \theta^2} \right)
\nonumber\\
-\frac{1}{2}\beta_1\frac{\partial^2 A_2}{\partial t^2}
+\gamma \left(| A_2|^2+ | A_1|^2\right)A_2=0. \nonumber
\end{eqnarray}
The nonlinear system of equations (\ref{NLC}) admits another infinite number of exact vortex solutions of kind of:

\begin{eqnarray}
\label{NSOL}
A_1=\sqrt{((n+1)^2-1)/\gamma}(r^{-1})\cos((n+1)\theta)\exp\left(i(\beta_1 z+\sqrt{2\alpha\delta^2}t)\right),\\
A_2=\sqrt{((n+1)^2-1)/\gamma}(r^{-1})\sin((n+1)\theta)\exp\left(i(\beta_1 z+\sqrt{2\alpha\delta^2}t)\right),
\end{eqnarray}
where $n=1,2...\infty$ is one infinite number. One important
consequence of these solutions is, that they are obtained as a
balance not only between the diffraction and the nonlinearity, but
mostly by the nonlinearity and the angular distribution. That is
why, we can expect one stable propagation of such vortices in z
direction. Right away we can point here, that conditions for
finiteness of the energy of the above solutions have been not found
up to now.

\section{ Pulses with equal transverse and longitudinal size $r_\perp\sim z_0$}

The investigation of the  propagation of optical pulses or modulated
waves with equal transverse and longitudinal size $r_{\perp}\sim
z_0$ is more convenient  to provide in coordinate system moving with
group velocity. That is why we apply a Galilean transformation to
the basic vector amplitude equation ($t'=t$, $z'=z-vt$), as
difference from the case of long pulses, where the "moving in time"
transformation is more natural. The equation (\ref{a11}) in this
case is transformed to:

\begin{eqnarray}
\label {a13} -i\left(\frac{\partial\vec A}{\partial t'}+
\left(n_2+\frac {k_0v}{2}\frac{\partial
n_2}{\partial\omega}\right)\left(\frac{\partial \left(\left|\vec
A\right|^2\vec A\right)}{\partial t'} -v\frac{\partial
\left(\left|\vec
A\right|^2\vec A\right)}{\partial z'}\right)\right)+\nonumber\\
\frac{v}{2k_0}\Delta_{\bot}\vec A -
\frac{v^3k_0^{"}}{2}\frac{\partial^2\vec A}{\partial z'^2} -
\frac{v}{2}\left(k_0^{"}+\frac{1}{k_0v^2}\right)
\left(\frac{\partial^2\vec A}{\partial t'^2}-2v\frac{\partial^2\vec
A}{\partial t'\partial z'}\right)+ \\
\frac{n_2 k_0 v}{2}\left|\vec A\right|^2\vec A=0, \nonumber
\end{eqnarray}
where $\Delta_{\bot}=\frac{\partial^2}{\partial x^2} +
\frac{\partial ^2}{\partial y^2}$. We use again the approximation
$\partial n_2/\partial\omega\simeq 0$. To estimate the influence of
the different terms on the propagation dynamics we rewrite the
equation (\ref{a13}) in dimensionless form. Defining the rescaled
variables:

\begin{eqnarray}
\label{rescal} \vec A=A_0\vec A';\ x=r_0x';\ y=r_oy';\ z"=r_0z';\
t"=t_0t',
\end{eqnarray}
 and constants:

\begin{eqnarray}
\label{norm5}
\alpha=2k_0r_0^2/t_0v=2k_0r_0;\  \beta=v^2k"k_0;\
\gamma=k_0^2r_0^2n_2\left|A_0\right|^2;\nonumber\\
\\
r_0/t_0=v;\ \gamma_1=2k_0r_0n_2\left|A_0\right|^2, \nonumber
\end{eqnarray}
equation (\ref{a13}) can be transformed in the following (the primes
and the seconds are not written for clarity again):

\begin{eqnarray}
\label {eq13}
 -i\left(\alpha\frac{\partial\vec A}{\partial t}+
\gamma_1\left(\frac{\partial \left(\left|\vec A\right|^2\vec
A\right)}{\partial t}-\frac{\partial \left(\left|\vec A\right|^2\vec
A\right)}{\partial z} \right)\right)+ \Delta_{\bot}\vec A
-\beta\frac{\partial^2\vec A}{\partial z^2}
-\nonumber\\
\\
 \left(\beta +1\right)\left(\frac{\partial^2\vec A}{\partial
t^2}-2\frac{\partial^2\vec A}{\partial t\partial z}\right) +
\gamma\left|\vec A\right|^2\vec A=0. \nonumber
\end{eqnarray}
We point out, that the normalizing is with equal longitudinal and
transverse dimensions  $r_0$. As it was pointed above, it is fulfilled usually for
few picosecond and femtosecond pulses of a  pulse with
transverse size of $1-5$ millimeters to hundred micrometers.

\subsection{Equation in optical region of media with week dispersion $\beta<<1$}

Dimensionless dispersion parameter $\beta$ in transparency region of
dielectrics and gases usually is of the order of $10^{-2}-10^{-3}$.
For example we calculate $\beta$ in the transparency region of pure
silica on $\lambda=1,55 \mu m $ using the standard Sellmeier
expression for dependence of the dielectric constant from frequency,
and the value really is of this order $ \beta=-0,0069 $. The
constant $\alpha$ has a value of $\alpha\approx 10^2$ ($\alpha
\approx 2r_0 k_0 $) if the slowly varying approximation is used. In
this case, when we investigate the optical pulses with power near to
the critical for self-focusing $\gamma\approx 1$, $\gamma_1\ll 1$,
our vector amplitude equation (\ref{eq13}) can be  transformed to
the next evolution equation:

\begin{eqnarray}
\label {eq14} -i\alpha\frac{\partial\vec A}{\partial t}+
\Delta_{\bot}\vec A - \left(\frac{\partial^2\vec A}{\partial
t^2}-2\frac{\partial^2\vec A}{\partial t\partial z}\right) +
\gamma\left|\vec A\right|^2\vec A=0.
\end{eqnarray}
The equation (\ref{eq14}) is one generalization of the well known
paraxial equation (\ref{parax}) in optical region for pulses with
equal transverse and longitudinal size.

\subsection{Equation in UV transparency region. Strong dispersion ($\beta\approx 1$)}

\begin{figure}[t]
\begin{center}
\includegraphics[width=130mm,height=85mm]{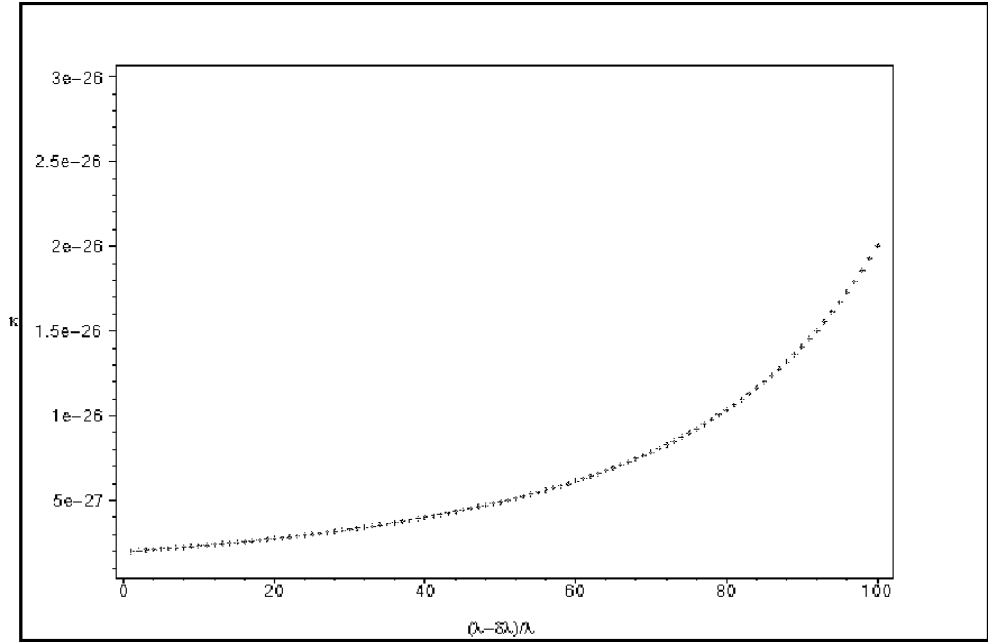}
\caption{Plot of the dispersion of the group velocity $k^"(\omega)$
in the  UV region from $\lambda=264$ nm to $\lambda=164$ nm for bulk
fused silica.}\label{Fig1}
\end{center}
\end{figure}

\begin{figure}[t]
\begin{center}
\includegraphics[width=130mm,height=85mm]{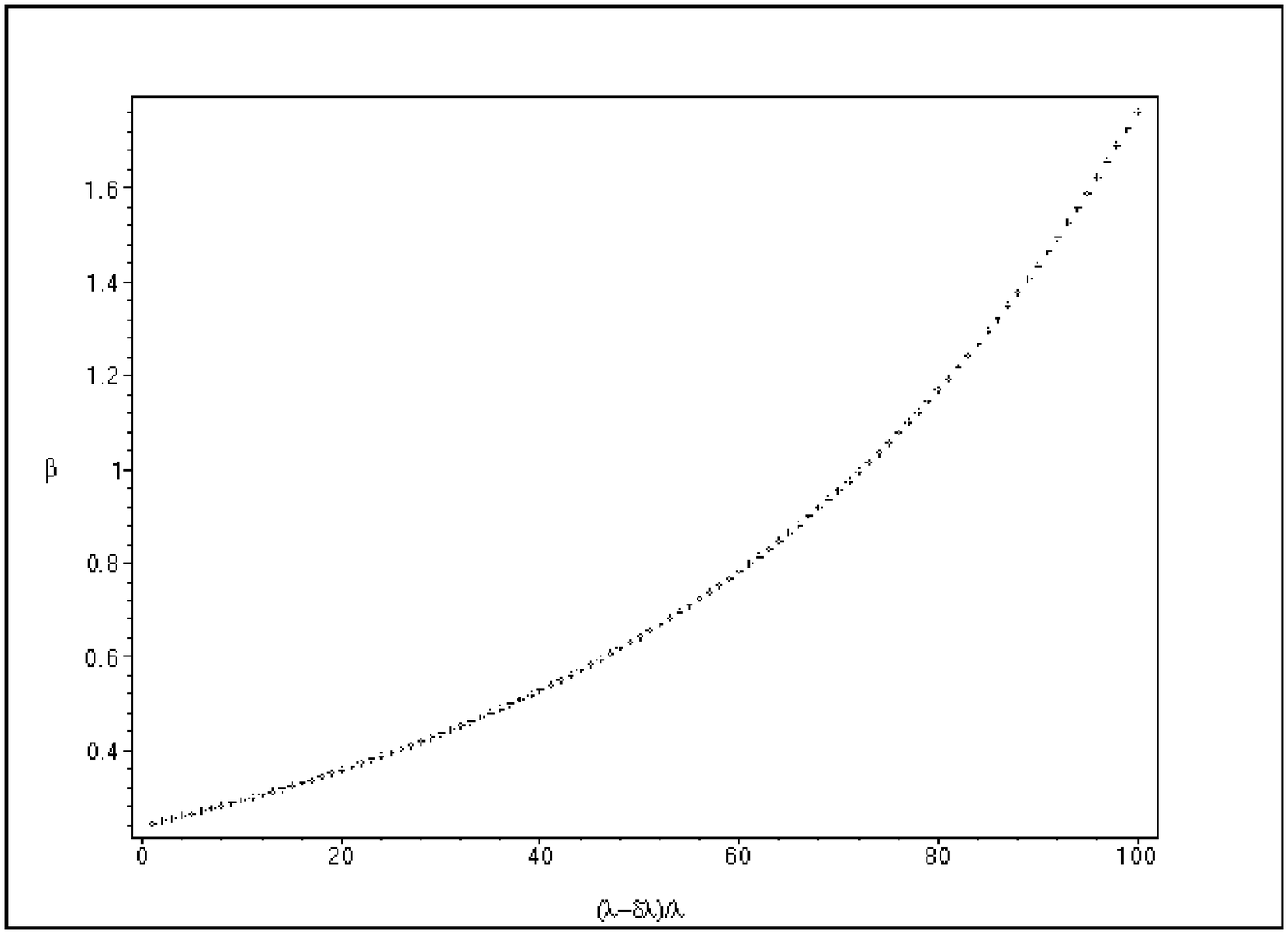}
\caption{ Plot of the dimensionless parameter $\beta=k_0v^2k^{"}$
within wavelengths from $\lambda=264$ nm to $\lambda=164$ nm for
bulk fused silica. The dispersion parameter reach a value $\beta\sim
1$ in UV region.}\label{Fig2}
\end{center}
\end{figure}

The critical dimensionless parameter $\beta$, which determines the
relation between dispersion and diffraction in the normalized
amplitude equation (\ref{eq13}), as it was established in the
previous paragraph, is very little in the optical region. In spite
of this, the temporal effects are presented in the corresponding
equation (\ref{eq14}) and for pulses with equal transverse and
longitudinal size can not be neglected.  The situation in UV region
is quite different. In this region we can reach $\beta\simeq 1$.
When we investigate optical pulses with power, near to the critical
for self-focusing, the corresponding nonlinear dimensionless
parameters admit values  $\gamma\approx 1$, $\gamma_1<<1$. Than the
equation (\ref{eq13}) is transformed to:

\begin{eqnarray}
\label {UV1} -i\alpha\frac{\partial\vec A}{\partial t}+
\Delta_{\bot}\vec A-\beta\frac{\partial^2\vec A}{\partial z^2}
-\left(\beta +1\right)\left(\frac{\partial^2\vec A}{\partial
t^2}-2\frac{\partial^2\vec A}{\partial t\partial z}\right) +
\gamma\left|\vec A\right|^2\vec A=0.
\end{eqnarray}
If we compare the equations in optical (\ref{eq14}) and UV region
(\ref{UV1}), it is seen that in UV region we can take in mind in
addition a negative second derivative in z direction of amplitude
function and we multiply the temporal effects of second order by
factor $\beta+1$. We use for example again  bulk fused  silica and
calculate the linear refractive index $n^2 (\omega) $ from the
Sellmeier relations. The investigation is provided for the spectral
region from wavelength $\lambda=0.264$ $\mu m$ to wavelength
$\lambda=0.164$ $\mu m$. The  group velocity $v_g$, dispersion
$k^"(\omega)$ and the dimensionless parameter $\beta(\omega)$ are
calculated. The typical characteristic of the laser pulse are
$t_0=5\times 10^{-12}$ $s$, group velocity $v_g\simeq1.8\times
10^{10}$ $cm/s$ and $r_{\perp}=z_0\simeq 1\times 10^{-1}$ $cm$. On
Figure \ref{Fig1}, Figure \ref{Fig2}  are presented correspondingly
the graphics of dispersion of the group velocity  $k^"(\lambda)$ and
dimensionless parameter $\beta(\lambda)$. As it is seen from Fig.2,
the dimensionless parameter $\beta$, really can reach values equal
to one in UV region for some dielectrics.

\vspace{3mm}

\subsection{Equation in media with strong negative dispersion \\ $\beta\approx -1$}

Now we come to the most interesting case in this investigation: the
possibility that the temporal effects of second order can be
neglected. As it can be traced out from basic amplitude equation
(\ref{a11}), and to see in the normalized equation (\ref{eq13}),
when $k_0v_g^2k"=\beta\cong-1$, the dispersion effects of second
order do not present in the corresponding amplitude equations and do
not influence on the propagation dynamics. Recently in \cite{KOV} it
was found that near the electronic resonances of gases and
dielectrics and also near the Langmuir frequency in cold plasma the
dispersion is negative, the dispersion parameter arises rapidly and
is of the order of $\beta\approx -1$. Again we investigate the
propagation of optical pulses with power near the critical  for
self-focusing ($\gamma\approx 1$, $\gamma_1<<1$). The amplitude
equation (\ref{eq13}) in this case can be written in the next simple
form of 3D+1 Vector Nonlinear Schr\"odinger equation (VNSE):

\begin{eqnarray}
\label{eq15}
-i\alpha\frac{\partial \vec A}{\partial t} +
\Delta_{\bot}\vec A + \frac{\partial^2\vec A}{\partial z^2} +
\left|\vec A\right|^2\vec A = 0.
\end{eqnarray}
Here we will study the possibility to obtain $\beta=-1$ in the next
three cases: 1. Near the isolated resonance of mercury vapor on
$\lambda=0.3653 \mu m $. 2. Near the Langmuir frequency in cold
electron plasma. 3. In deep UV and R\"o region for dielectrics,
gazes and  metals.

\subsubsection{Dispersion parameter $\beta$ near the isolated electron
resonance}

When $\varepsilon(\omega)\sim
1/\omega$ the wave vector can be written as:

\begin{eqnarray}
\label{eps}
k(\omega)=\frac{\sqrt{\omega}}{c}.
\end{eqnarray}
Calculating $\beta=v^2k"k$ for dispersion relation of kind of
(\ref{eps}), right away we obtain $\beta=-1$. This gives us a
confidence to look for such zones near the electronic resonances
where $\varepsilon(\omega)\sim 1/\omega$. Near to some of the
isolated electron resonances of metal vapors the dielectric constant
can be expressed by \cite{BOYD}:

\begin{eqnarray}
\label{res} \varepsilon(\omega)=1+\frac{f4\pi Ne^2}{m\omega_0}
\frac{\omega_0-\omega}{\left(\omega_0-\omega\right)^2+\gamma^2}.
\end{eqnarray}
The investigation are provided for resonance frequency of mercury
vapor on $\omega_0=2\pi c/\lambda_0=0.515568.10^{16}$ Hz, oscillator
strength $f=0.3$, and natural (i.e. radiative) linewidth
$\gamma=1/4,53\lambda_0^2=1,654.10^8$ Hz. We use Lorenz shape of the
line, as we investigate the rarefied vapor and looking in zones
outlying from resonance on distance more than $4\gamma$, where the
natural shape of the line dominated. In Figure 3 it is shown the
form of the dielectric constant for number density $N=1,2.10^{15}$.
In the Figure 4 it is presented the dimensionless dispersion
parameter $\beta$. As it is well seen on the wing on the resonance
on distance about $4\gamma$ we have spectral zone where $\beta\simeq
-1$. Right away it must be pointed that $\beta$ depends strongly on
the number density N. The number density of the order of $N\approx
1.10^{13}-1.10^{15}$ appears to be most suitable for large zones
with $\beta\sim-1$. Seems, for such number density the wing of the
resonance in best way can be approximated with
$\varepsilon(\omega)\sim 1/\omega$.

\begin{figure}[t]
\begin{center}
\includegraphics[width=130mm,height=85mm]{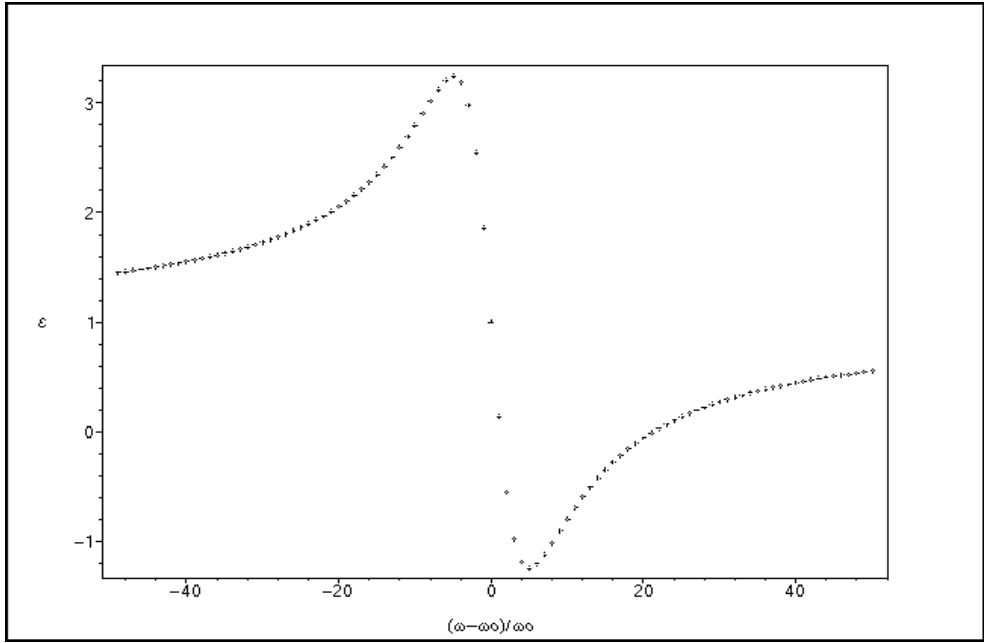}
\caption{ Frequency dependence of the dielectric constant  for
resonance frequency of mercury vapor of $\omega_0=2\pi
c/\lambda_0=0.515568.10^{16}$ Hz ($\lambda=365$ nm), and natural
linewidth $\gamma=1/4,53\lambda_0^2=1,654.10^8$ Hz.}\label{Fig3}
\end{center}
\end{figure}
\vspace{3mm}

\begin{figure}[t]
\begin{center}
\includegraphics[width=130mm,height=85mm]{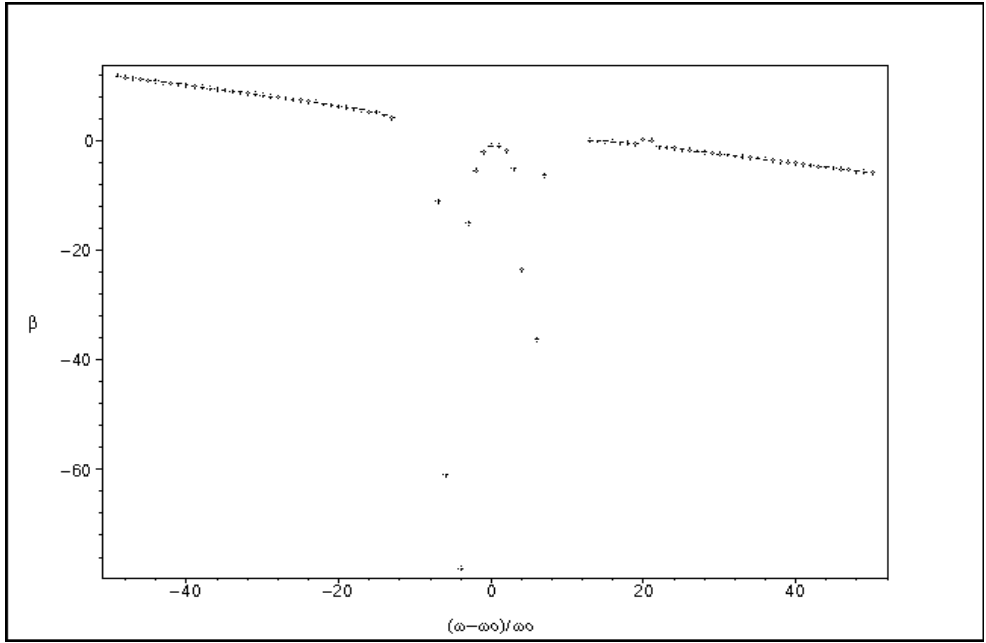}
\caption{ Plot of the dimensionless parameter $\beta=k_0v^2k^{"}$
near to resonance frequency of mercury vapor on $\omega_0=2\pi
c/\lambda_0=0.515568.10^{16}$ Hz. On distance $4\gamma$ we find the
spectral zone where $\beta\simeq -1$.}\label{Fig4}
\end{center}
\end{figure}

\subsubsection{Dispersion parameter $\beta$ near to Langmuir
frequency in cold plasma}

Dielectric constant  in the case of cold electron plasma is given by the expression \cite{KARP}:

\begin{eqnarray}
\label{plazma}\varepsilon(\omega)=1-\frac{\omega_p^2}{\omega^2},
\end{eqnarray}
where $\omega_p=\sqrt{4\pi N_1e^2/m}$ is the plasma frequency and
$N_1$ is the number of free electron particles per $cm^{-3}$. Here
we investigate a typical number which characterized the plasma
density in generators used for nuclear fusion. In Figure 5 is
presented $\varepsilon(\omega)$ for frequencies higher and close to
plasma frequency and number density $N_1\cong 1.10^{17}$ $cm^{-3}$.
The frequency is normalizing on spectral width of optical pulse
$\delta\omega=1.10^{10}$ Hz, which corresponds to pulse with time
duration $\Delta t\simeq 2.10^{-10}$ seconds. The corresponding to
this dielectric constant dispersion parameter $\beta(\omega)$ is
presented in Figure 6. As it is seen from the graphics, the large
spectral zone exists, more than twenty spectral widths from plasma
frequency, where $\beta\cong -1$. The transverse and longitudinal
size such optical pulse valued from $0.5-1$ $cm$.  An important
relation is pointed out with investigation of dependence of $\beta$
from time duration of the pulses and number density. With decreasing
the time and spatial duration of the pulses (fs pulses) and
increasing the number density, this large spectral zone with
$\beta\cong -1$  keeps on.

\begin{figure}[t]
\begin{center}
\includegraphics[width=130mm,height=85mm]{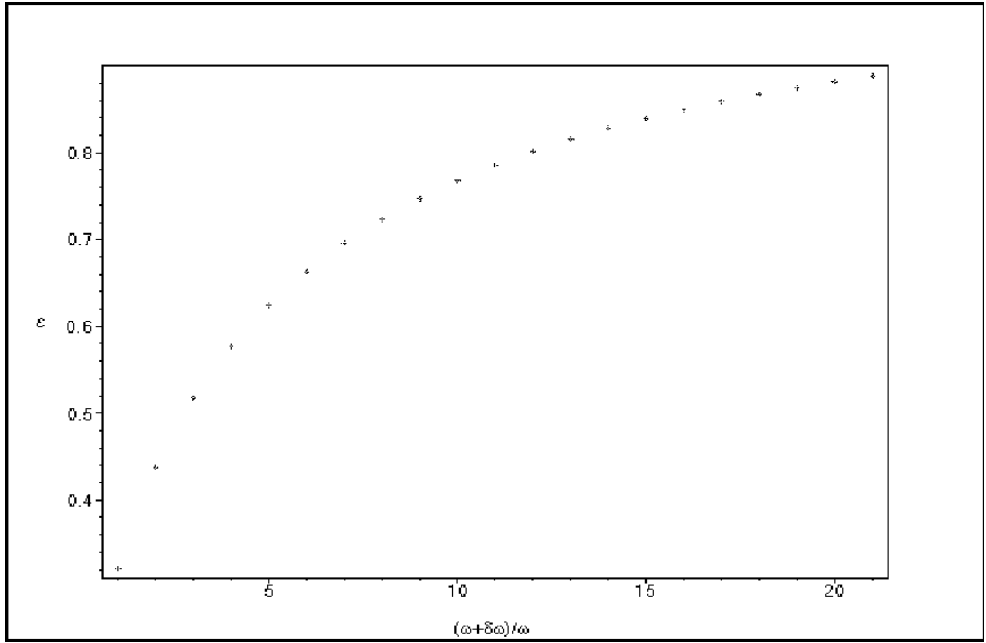}
\caption{Frequency dependence of the dielectric constant near to
Langmuir frequency in cold plasma for number density $N_1\sim
1.10^{17}$. The frequency is normalizing on spectral width of the
optical pulse $\delta\omega=1.10^{10}$ Hz.}
\end{center}
\end{figure}

\begin{figure}[t]
\begin{center}
\includegraphics[width=130mm,height=85mm]{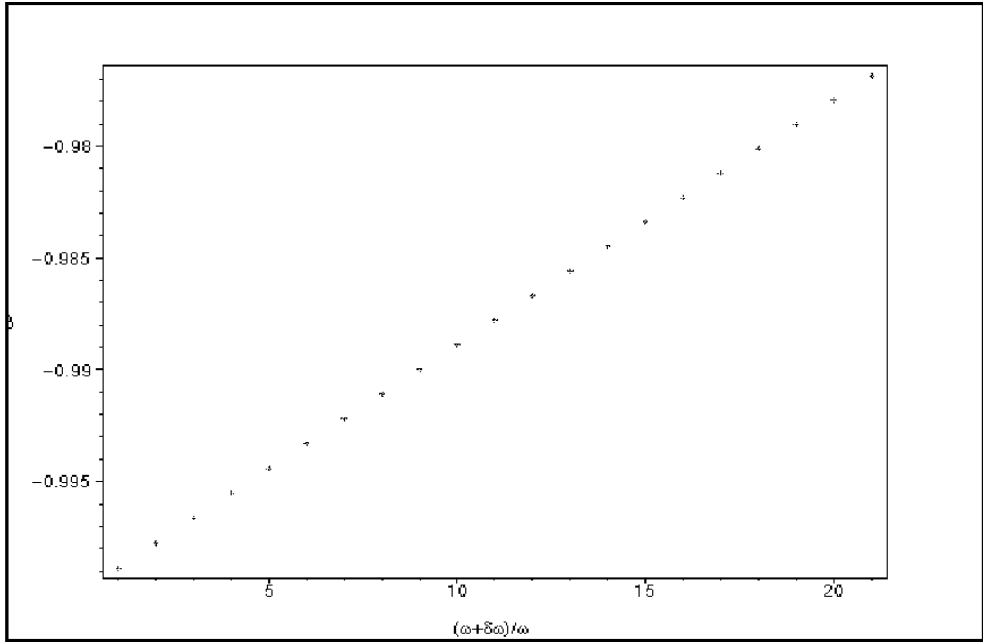}
\caption{ The corresponding dimensionless dispersion parameter
$\beta$ near the plasma frequency. Large spectral sone with
$\beta\simeq -1$ is established.}\label{Fig6}
\end{center}
\end{figure}

\subsubsection{Dielectric constant $\varepsilon(\omega)$ and dispersion parameter $\beta$ from  deep UV to
R\"o region}

As it was established in \cite{LAN}, it is possible to find equal
dispersion relation of the dielectric constant $\varepsilon(\omega)$
for different kind of materials, as dielectrics, gases,
semiconductors, metals, when the carrying frequency of optical wave
is much higher than the resonances of this media. In these cases the
localized energy of optical wave is much higher than the critical
for ionization of the materials and the linear and nonlinear
response is the same as in cold plasma. Practically the electrons
interact as free electrons on these frequencies. It is not hard to
find that in this region again we have the dispersion relation of
dielectric constant of the kind of:

\begin{eqnarray}
\label{plazma1}\varepsilon(\omega)=1-\frac{\omega_p^2}{\omega^2}.
\end{eqnarray}

The critical parameter which characterized the plasma frequency in
these materials is the number density $N_1$ of the free electrons.
The typical number for semiconductors is $N_1\cong 1.10^{18}$ and
for metals is $N_1\cong 1.10^{22}$. That is why, the application of
this formula started from deep UV for light elements as H, Li, e.c.
and from R\"o frequencies in heavy elements. For optical waves when
$\omega>\omega_{p}$, there are no difference between metals,
semi-conductors, dielectrics and metals. All of them are transparent
for optical waves with carrying frequency $\omega_0>\omega_{p}$.
Again we can obtain, that near the corresponding plasma frequency of
these materials, large frequency zone with negative dispersion
parameter of order of $\beta\sim -1$ exists, if the picosecond and
femtosecond pulses are used. For all of these three important cases
the propagation of localized optical pulses is governed by the
vector 3D+1 Nonlinear Schr\"odinger Equation (\ref{eq15}). The
numerical experiment provided in \cite{KOV}  show a stable
propagation of the vortex solutions. The intensity picture of these
multi-component solutions is truly symmetric and look like as stable
Fraunhofer distribution. The internal structure of these vortices is
presented by the spherical harmonics with integer number.

\section{Exact vortex solutions on VNSE}

The solutions of the 3D+1 Vector Nonlinear Schr\"odinger amplitude
equation (\ref{eq15}) in a fixed basis are \cite{KOV}:

\begin{eqnarray}
\label{c1}
A_x = \sqrt 2\frac{\exp\left({i\sqrt{\alpha\Omega}
r}\right)}{r} \sin \theta \cos \varphi \exp \left(i\Omega t
\right),
\end{eqnarray}

\begin{eqnarray}
\label{c2} A_y = \sqrt 2\frac{\exp\left({i\sqrt{\alpha\Omega}
r}\right)}{r} \sin \theta \sin \varphi \exp \left(i\Omega t
\right),
\end{eqnarray}

\begin{eqnarray}
\label{c3} A_z = \sqrt 2\frac{\exp\left({i\sqrt{\alpha\Omega}
r}\right)}{r} \cos \theta \exp \left(i\Omega t \right),
\end{eqnarray}
where $r=\sqrt{x^2 + y^2 + z^2}$, $\theta= \arccos\frac zr$ and $\varphi=arctan\frac xy $
are the moving spherical variables of the independent variables
$x, y, z$. The corresponding real solutions  $\Re(\vec
{A})$, are:

\begin{eqnarray}
\label{eq30} \Re(A_x)=\frac{1}{2i}\left( A_x-A_x^{*}\right)=
\sqrt{2}\frac{\sin\left(\sqrt{\alpha\Omega}r+\Omega t\right)}{r}\sin
\theta \cos \varphi ,
\end{eqnarray}

\begin{eqnarray}
\label{eq31} \Re(A_y)=\frac{1}{2i}\left( A_y-A_y^*\right)= \sqrt {2}
\frac{\sin\left({\sqrt{\alpha\Omega} r+\Omega t}\right)}{r}\sin
\theta \sin \varphi ,
\end{eqnarray}

\begin{eqnarray}
\label{eq32} \Re(A_z)=\frac{1}{2i}\left( A_z-A_z^*\right)= \sqrt {2}
\frac{\sin\left({\sqrt{\alpha\Omega} r+\Omega t}\right)}{r}\cos
\theta.
\end{eqnarray}
It is important to point, that in \cite{KOV} conditions for
finiteness of the energy of the solutions (\ref{eq30})-(\ref{eq32})
was found. As it can be established, the averaged with one time
period intensity field of these solutions admits total radial
symmetry:

\begin{eqnarray}
\label{int} \langle A_x\rangle^2+\langle A_y\rangle^2+\langle
A_z\rangle^2= 2\frac{\sin^2\left(\sqrt{\alpha\Omega}r\right)}{r^2}.
\end{eqnarray}
Thus, in one experiment the solution will look like as  Fraunhofer
distribution. Figure 7 presents the surface $\langle\vec
A\rangle^2_{z=0}$ of the solutions (\ref{eq30})-(\ref{eq32}).

\begin{figure}[t]
\begin{center}
\includegraphics[width=65mm,height=55mm]{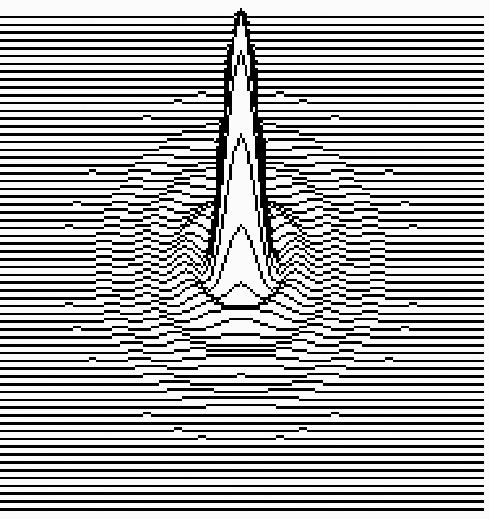}
\caption{ Fraunhofer distribution for  the averaged with one time
period intensity of the solutions (\ref{int}) of 3D+1 Vector
Nonlinear Schr\"odinger equation (\ref{eq15}).}\label{Fig7}
\end{center}
\end{figure}
We can recognize the vortex structures of these solutions only if we
see the field distribution and the intensity of one of the
component, for example $\langle A_x\rangle$. On Figure 8 we present
the surface of one of the components $\langle A_{x,|z=0}\rangle$,
rewritten in Cartesian coordinates:

\begin{eqnarray}
\label{komp} \langle
A_x\rangle=\sqrt{2}\frac{x\sin\left(\sqrt{x^2+y^2+z^2}\right)}
{x^2+y^2+z^2}.
\end{eqnarray}
It is clearly seen, that the surface of our three dimensional
solutions possess linear phase dislocations in space, which
characterizes the vortex structures \cite{NB}.

\begin{figure}[t]
\begin{center}
\includegraphics[width=65mm,height=55mm]{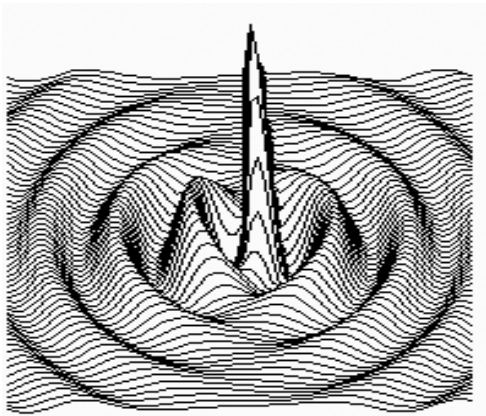}
\caption{Surface of the one of the component $\langle
A_{x,|z=0}\rangle$ of the averaged with one time period vortex
solutions (\ref{komp}). It is clearly seen linear phase dislocation
for these solutions.}\label{Fig8}
\end{center}
\end{figure}
The intensity of this component $\langle A_x^2\rangle$, which is one
$l=1, m=1$ distribution, is presented in Figure 9.  This picture is
possible to see in one experiment, when an optical vortex of kind
(\ref{eq30})-(\ref{eq32}) is passed trough the linear polarizer.
Then the internal structure of these vortex solutions composed by
spherical harmonics with $l=1, m=\pm1, 0$ distribution, will be
seen. Our solutions (\ref{eq30})-(\ref{eq32}) admits total angular
momentum $l=1$ and line phase singularities for the different
components. That is why, in spite of the fact that the intensity
field is total symmetric, we recognize the internal vortex
structures of these solutions.

\begin{figure}[t]
\begin{center}
\includegraphics[width=65mm,height=55mm]{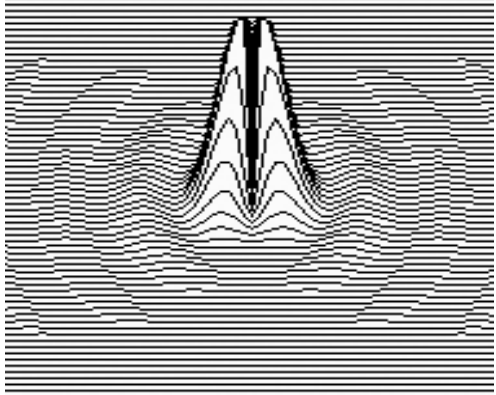}
\caption{The intensity profile of one of the components $\langle
A_{x,|z=0}\rangle^2$, which corresponds to $l=1,m=1$
distribution.}\label{Fig9}
\end{center}
\end{figure}

\section {Conclusion}

In conclusion, we investigated the application of the nonlinear
amplitude equation (\ref{a11}), in different spectral zones,
starting from the optical region, and extending up to the R\"o
region of dielectrics. Different generalizations of the paraxial
equation are obtained, depending from the spectral zones, the
material constants, and the initial shape of the pulses (long pulses
or optical bullets). We find large spectral region near the plasma
frequency in cold plasma for fs and few picosecond optical pulses,
where the temporal effects of second order in the amplitude
equations (\ref{a11}) can be neglected. The same result is obtained
also for spectral regions on the 'wings' of the electron resonances
in metal vapors. In the last case the conditions to obtain such
effects depend strongly on the number density of the gases. In these
spectral zones the propagation of optical pulses is governed by the
amplitude vector nonlinear Schr\"odinger equation (\ref{eq15}) and,
as it was predicted in \cite{KOV}, stable 3D optical vortices can be
observed. They look like as   Fraunhofer distribution in the space.
After passing trough the linear polarizer the vortex structures will
be seen. The vector theory presented in this paper can be extended
also in respect to different kind of nonlinearities and
polarizations of the optical field.

\section {Appendix 1. Nonlinear polarization of two and three components of the same frequency}

The electric field associated with linear or elliptically polarized
optical wave can be written in the form

\begin{eqnarray}
\vec{E}(x,y,z,t)=\frac{1}{2i}\left(\left(\vec{x}A_x+\vec{y}A_y\right)\exp(i\omega_0t)
-c.c.\right),
\end{eqnarray}
where $A_x$ and $A_y$ are the complex amplitudes of the polarization
components of a wave with the frequency $\omega_0$.

\subsection{Linearly polarized components}
Let investigate the polarization dynamics of two initially linearly
polarized components of the electrical field. Then, there is no
initial phase different between the components, and the complex
amplitudes can be expressed as a product of two real amplitude
functions with equal phases: $ \Re(\hat{A}_x)\exp(i\phi)$, $
\Re(\hat{A}_y)\exp(i\phi) $. Calculating the relation between the
linear polarized components in the same manner, as it was performed
for plane wave \cite{YR}, we obtain the same kind  relation for the
components of the electrical field:

\begin{eqnarray}
\label{YAR}
 \frac{E_y}{E_x}=\frac{\Re(\hat{A}_y)}{\Re(\hat{A}_x)},
\end{eqnarray}
where $\Re(\hat{A}_y)$  and $\Re(\hat{A}_x)$ are real functions, not
real constants as in the case of plane wave. We investigate the
propagation dynamics of optical pulses with arbitrary localized
amplitudes. The soliton case appear as partial case, when equal
amount of nonlinear and linear phase shift are compensated, the
initial phase different do not changed, and the linear polarization
keeps on. For optical pulses with arbitrary amplitudes, as the media
is isotropic, the diffraction and the dispersion add equal phases to
the two components and they can be written generally in the form:

\begin{eqnarray}
\label{DIFDIS1}
A_x=\Re(A_x)\exp(i(K_{diff}z+\Omega_{disp}t)+\phi)=\Re(A_x)\exp(i\Phi(z,t)),
\end{eqnarray}
\begin{eqnarray}
\label{DIFDIS2}
A_y=\Re(A_y)\exp(i(K_{diff}z+\Omega_{disp}t)+\phi)=\Re(A_y)\exp(i\Phi(z,t)),
\end{eqnarray}
where $K_{diff}$ is the phase shift due to diffraction,
$\Omega_{disp}$ is the phase shift due to dispersion, $\Phi(z,t)$ is
the equal phase components for the both amplitudes, and $\Re(A_y)$,
$\Re(A_y)$ are the real part of solutions for the amplitudes on
distance z in isotropic media. For arbitrary solutions of kind
(\ref{DIFDIS1}), (\ref{DIFDIS2}) the relation (\ref{YAR}) keeps on
for the real part of the amplitudes. It is following from the fact,
that the diffraction and the dispersion effects do not change the
initial linear polarization of the pulses (do not yield linear
birefringence in isotropic materials) . The nonlinear polarization
$\vec{P}_{nl}$ in the case of isotropic medium is well known and can
be written in the form \cite{MT,LAN}:

\begin{eqnarray}
\label{PNL}
\vec{P}_{nl}=A(\omega)\left(\vec{E}\cdot\vec{E^*}\right)\vec{E}+
\frac{B(\omega)}{2}\left(\vec{E}\cdot\vec{E}\right)\vec{E^*}.
\end{eqnarray}
Substituting the expression for the two linearly polarized component
(\ref{DIFDIS1}) and (\ref{DIFDIS2}) in (\ref{PNL}), right way we
obtain for the nonlinear polarization of the different components
\cite{LAN}:

\begin{eqnarray}
\label{PNL1}
P^x_{nl}=\left(A(\omega)+\frac{B(\omega)}{2}\right)\left(|A_x|^2+|A_y|^2\right)A_x,
\end{eqnarray}
\begin{eqnarray}
\label{PNL2}
P^y_{nl}=\left(A(\omega)+\frac{B(\omega)}{2}\right)\left(|A_x|^2+|A_y|^2\right)A_y.
\end{eqnarray}
We point here, that the expression is valid not only for nonlinear
nonresonant optical response, when $A=B=const$, but also for
nonlinearity near to some of the resonances, when $A(\omega)\neq
B(\omega)$. In the same way we can obtain, that the components of
the nonlinear polarization for one linear polarized vector in 3D
space take forms:

\begin{eqnarray}
\label{PNL3}
P^x_{nl}=\left(A(\omega)+\frac{B(\omega)}{2}\right)\left(|A_x|^2+|A_y|^2+|A_z|^2\right)A_x,\\
P^y_{nl}=\left(A(\omega)+\frac{B(\omega)}{2}\right)\left(|A_x|^2+|A_y|^2+|A_z|^2\right)A_y,\\
P^z_{nl}=\left(A(\omega)+\frac{B(\omega)}{2}\right)\left(|A_x|^2+|A_y|^2+|A_z|^2\right)A_z.
\end{eqnarray}

\subsection{Elliptically polarized components}

In the case of elliptically polarized light the relation (\ref{YAR})
take form:
\begin{eqnarray}
\label{YAR1}
\frac{E_y}{E_x}=\frac{\Re(A_y)}{\Re(A_x)}\exp(i(\delta_y-\delta_x)),
 \end{eqnarray}
where $\delta_y-\delta_x\neq 0,\ n\pi,\ \pi/2, $ is the different of
the phase functions for the components of the vector field. Since
the diffraction and the dispersion add equal amount of phase shift
to the different components in isotropic media, the phase different
keeps on again. If we substitute $E_x$ and $E_y$ with such phase
difference in the expression for nonlinear polarization (\ref{PNL}),
and take in mind that $A=B=const$ for case of nonresonant electronic
nonlinearities, the next expression for the components can be found
\cite{AGR}:

\begin{eqnarray}
\label{PNLX}
P^x_{nl}=const\left[\left(|A_x|^2+\frac{2}{3}|A_y|^2\right)A_x+\frac{1}{3}(A^*_xA_y)A_y\exp(2i(\delta_x-\delta_y))\right],
\end{eqnarray}

\begin{eqnarray}
\label{PNLY}
P^y_{nl}=const\left[\left(|A_y|^2+\frac{2}{3}|A_x|^2\right)A_y+\frac{1}{3}(A^*_yA_x)A_x\exp(-2i(\delta_x-\delta_y))\right].
\end{eqnarray}
The last terms in equations (\ref{PNLX}) and (\ref{PNLY}) present
degenerate four wave mixing process ($\omega_1=\omega_2=\omega_0$).
Since $\delta_x-\delta_y\neq 0$, this leads to periodically exchange
of energy between the elliptically polarized components. Thus, such
one exchange of energy can be observed in isotropic material as
gases, metal vapors and crystals. When the phase different vanish
($\delta_x-\delta_y= 0$) in the case of linearly polarized
components, the amplitude functions admit again equal phases, and
the nonlinear polarization (\ref{PNLX}) and (\ref{PNLY}) can be
transformed to the nonlinear polarization of linearly polarized
components (\ref{PNL1}) and (\ref{PNL2}). This is one generalization
of the results obtained for parametric solitons in $\chi^{(3)}$
media \cite{PARKOV}. When there is no phase different between the
components, the parametric therm appear as usual cross-phase term in
the dispersion relations, and can be take in mind to determinate the
right amplitude constants. The polarization dynamics in materials
which admit modal birefringece is more complicate. In single-mode
fiber, where the linear birefgingent effect appear \cite{KAM},
additional polarization mode dispersion exists \cite{MEN}. This lead
to random birefringence in optical fibers and random phase different
in the last parametric terms in (\ref{PNLX}) and (\ref{PNLY}). Our
numerical experiment show that the parametric effects with random
phases are not effective, and can be neglected. One transitions to
the Manakov system with coefficients equal to one in front of the
cross-terms \cite{MAN} is possible again for birefringent materials
at elliptically angle $35^\circ$ between the main axes \cite{MEN1}.

\end{document}